\documentclass[11pt]{article}
\usepackage[top=1in,bottom=1in,left=1in,right=1in]{geometry}
\usepackage[onehalfspacing]{setspace}
\usepackage{authblk}
\usepackage{amsfonts,amssymb,amsmath,amsthm}
\usepackage{xcolor}
\usepackage{natbib}
\usepackage{hyperref}
\hypersetup{pdfborder={0 0 0},colorlinks=true,allcolors=blue,
  pdftitle={Regression Discontinuity Approaches to Evaluate Threshold-Guided Interventions in Observational Studies},
  pdfauthor={Matias D. Cattaneo and Rocio Titiunik}}
\begin{document}
\title{Regression Discontinuity Approaches to Evaluate Threshold-Guided Interventions in Observational Studies}
\author[1]{Matias D. Cattaneo}
\author[2]{Roc\'{i}o Titiunik}
\affil[1]{Department of Economics, Princeton University, Princeton, New Jersey.}
\affil[2]{Department of Politics, Princeton University, Princeton, New Jersey.}
\date{}
\maketitle

\begin{center}
\small Forthcoming in \emph{Journal of the American Medical Association}.\\
Published online November 2, 2026. \href{https://doi.org/10.1001/jama.2026.22316}{doi:10.1001/jama.2026.22316}
\end{center}

When using observational data to learn about the effect of an intervention, unmeasured differences between patients who receive different interventions can confound estimates of treatment effects \citep{Rosenbaum2010-book}. Regression discontinuity (RD) approaches can address this problem when treatment (eg, intervention, program, or policy) depends on whether a patient measurement---such as a laboratory test result---crosses a prespecified threshold \citep{Maciejewski-Basu_2020_JAMA}. This measurement is often called the score or running variable in RD analyses. Threshold-based medical guidelines therefore create opportunities to use RD approaches in clinical research \citep{Cattaneo-Keele-Titiunik_2023_SIM}.

Goulden and colleagues used an RD analysis to study the effect of contrast-enhanced computed tomographic pulmonary angiography (CTPA) on kidney function among patients undergoing D-dimer testing for suspected pulmonary embolism \citep{GouldenEtal2021-JAMA}. Typical clinical guidance at the time recommended CTPA when the D-dimer level was above 500 ng/mL, with some hospitals using a 460- or 470-ng/mL cutoff. The analysis compared patients with D-dimer levels just above and below the cutoff, using estimated glomerular filtration rate (eGFR) as the measure of kidney function.

\section*{Why Are RD Approaches Used for Identifying Causal Effects in Observational Studies?}

When treatment assignments are not randomized, patient groups receiving different treatments often differ in terms of both measured and unmeasured characteristics that influence the outcome. Adjusting only for measured characteristics may not control for confounding by unmeasured characteristics. When treatment eligibility or treatment recommendation changes at a cutoff, patients with scores just above and below the cutoff are often likely to be similar in both their measured and unmeasured baseline characteristics \citep{Lee-Lemieux_2010_JEL}. An RD analysis compares those 2 groups to obtain a valid estimate of the treatment effect.

\section*{How Does an RD Approach Work and How Is It Evaluated?}

The key assumption that justifies an RD analysis is the comparability of patients with measurements just below and above the cutoff. This assumption can be formalized in 2 ways: a continuity framework or a local randomization framework. Under the continuity framework, the average outcomes patients would experience under each intervention are assumed not to change abruptly at the cutoff, and therefore a jump in observed average outcomes at the cutoff can be attributed to the change in treatment \citep{Cattaneo-Idrobo-Titiunik_2020_CUP}. Under the local randomization framework, whether a patient's score falls above or below the cutoff is assumed to be randomly assigned within a small region around the cutoff, permitting the use of analytic methods similar to those for randomized trials \citep{Cattaneo-Idrobo-Titiunik_2024_CUP}.

An RD approach is called \emph{sharp} when every patient receives the intervention indicated by the threshold rule and \emph{fuzzy} when adherence to the rule is imperfect. In a sharp RD approach, the causal effect of interest is the average treatment effect for patients at or near the cutoff because there is no difference between the rule-defined treatment and the treatment received \citep{Cattaneo-Idrobo-Titiunik_2020_CUP}.

In contrast, a fuzzy RD approach yields 3 related causal effects. The first is the effect of the threshold rule on the probability of receiving the intervention, called the \emph{first-stage effect} or \emph{take-up effect}. This effect is a measure of the concordance---or lack thereof---between actual care and care dictated by the rule. The second causal effect is the effect of the threshold rule on the outcome, analogous to the intention-to-treat effect of randomized treatment assignment on the outcome in a randomized clinical trial \citep{Detry-Lewis_2014_JAMA}. The third causal effect is the effect of receiving the intervention among the complier subgroup, namely those patients whose intervention was changed because their score fell above rather than below the cutoff. Estimating this effect, known as the \emph{complier average treatment effect}, requires making additional assumptions for causal interpretation \citep{Cattaneo-Idrobo-Titiunik_2024_CUP}.

An RD analysis requires data on the score and outcome, knowledge of the cutoff, and, for a fuzzy RD analysis, data on the intervention actually received. Baseline characteristics are also useful for evaluating the key RD assumption of comparability near the cutoff. The analysis focuses on patients within a small window around the cutoff. Under the continuity framework, separate local linear regressions of the effect of the score on the outcome are fitted above and below the cutoff, with the score centered at the cutoff. Each intercept estimates the average outcome at the cutoff on its side, and the difference between the intercepts is the estimated average effect of treatment assignment on the outcome. Under the local randomization framework, the effect of the threshold rule is commonly estimated by comparing average outcomes just above and below the cutoff within the chosen window. For a fuzzy RD analysis, under either framework, the complier average treatment effect is estimated by dividing the effect of the threshold rule on the outcome by its effect on the probability of receiving the intervention \citep{Cattaneo-Idrobo-Titiunik_2020_CUP,Cattaneo-Idrobo-Titiunik_2024_CUP}.

To evaluate an RD analysis, investigators can use falsification checks that can reveal patterns inconsistent with the key RD assumption of comparability near the cutoff. The main check, analogous to evaluating covariate balance in a randomized clinical trial, is whether important baseline characteristics show a discontinuity (or jump) at the cutoff because, in a valid RD analysis, they should not. Other checks examine whether the distribution of scores jumps at the cutoff, conclusions are similar across reasonable window choices around the cutoff, apparent effects occur at artificial cutoffs where the treatment rule does not change, and results change after excluding observations closest to the cutoff. These checks cannot prove that an RD analysis is valid, but they can reveal evidence against its assumptions \citep{Cattaneo-Idrobo-Titiunik_2020_CUP}.

\section*{What Are the Limitations of RD Analysis?}

An RD analysis requires a known threshold rule that changes eligibility for, recommendation of, or receipt of an intervention. Although RD analyses can be extended to more complex settings, including those involving multiple cutoffs \citep{Cattaneo-Titiunik_2022_ARE}, many clinical questions do not arise from threshold-based decisions. Randomized clinical trials or other observational approaches must be used in those settings.

Even when an RD approach is applicable, causal interpretation may fail if patients or clinicians can influence the score, its timing, or how it is measured or recorded to obtain a preferred intervention because patients just above and below the cutoff may then differ systematically. Interpretation is also difficult when another aspect of care or policy changes at the same cutoff because any jump in the outcome cannot be uniquely attributed to the intervention of interest.

RD estimates rely primarily on observations close to the cutoff. Consequently, even a large study may yield imprecise estimates if few patients lie near the cutoff or, in a fuzzy RD analysis, if crossing the threshold only slightly changes the probability of receiving the intervention. Moreover, the estimated effects are local: they apply to patients at or near the cutoff and may not generalize to patients with substantially different scores. Extrapolation beyond the cutoff requires making additional strong assumptions.

\section*{How Did Goulden and Colleagues Apply RD?}

Goulden and colleagues studied 156\,028 emergency department patients who underwent D-dimer testing for suspected pulmonary embolism. Patients with D-dimer levels above the applicable cutoff were much more likely to receive contrast-enhanced CTPA than those below it, but adherence to this cutoff was imperfect, so the authors used a fuzzy RD analysis \citep{GouldenEtal2021-JAMA}.

The study illustrates how comparisons near the cutoff can reduce confounding. Across the full sample, patients whose D-dimer levels were above the cutoff were, on average, approximately aged 60 years, compared with 50 years among those whose levels were below the cutoff, a difference that could confound comparisons of kidney function. Near the cutoff, the age difference was much smaller, and other measured baseline characteristics were also similar \citep{GouldenEtal2021-JAMA}. Manipulation of the score was unlikely because patients could not control their measured D-dimer levels.

Using local linear regressions on either side of the cutoff, the investigators estimated (1) the effect of crossing the D-dimer cutoff on long-term eGFR and (2) the effect of receiving contrast-enhanced CTPA among patients whose receipt of CTPA was induced by the threshold rule---the complier average treatment effect.

\section*{How Should the Goulden Study Be Interpreted?}

The estimated effect of crossing the D-dimer cutoff on long-term eGFR was $-0.1$ mL/min/1.73~m$^2$ (95\% CI, $-1.2$ to $1.1$). The estimated complier average treatment effect of receiving contrast-enhanced CTPA was $-0.4$ mL/min/1.73~m$^2$ (95\% CI, $-4.9$ to $4.0$) \citep{GouldenEtal2021-JAMA}. Both estimates were close to 0 and both 95\% CIs include no effect. Assuming the RD and fuzzy RD assumptions hold, these results provide no evidence that contrast-enhanced CTPA harms long-term kidney function among patients with D-dimer levels near the cutoff whose receipt of CTPA was affected by the threshold rule. They do not establish the absence of harm for all patients: the 95\% CIs for the complier effect allow for both decreases and increases in eGFR, and the estimate may not generalize to patients with substantially different D-dimer levels or those whose receipt of CTPA would not change at the cutoff.

\section*{Article Information}

\noindent\textbf{Corresponding Author:} Roc\'{i}o Titiunik, PhD, Department of Politics, Princeton University, Fisher Hall No.\ 001, Princeton, NJ 08544 (\href{mailto:titiunik@princeton.edu}{titiunik@princeton.edu}).

\noindent\textbf{Conflict of Interest Disclosures:} None reported.

\noindent\textbf{Funding/Support:} The authors gratefully acknowledge financial support from the National Science Foundation (SES-2019432, SES-2241575).

\noindent\textbf{Role of the Funder/Sponsor:} The National Science Foundation had no role in the preparation, review, or approval of the manuscript or decision to submit the manuscript for publication.

\bibliography{CT_2026_JAMA}
\bibliographystyle{jasa}
\end{document}